\documentclass[a4,12pt]{article}

\pdfoutput=1

\usepackage{amsmath}
\usepackage{amsfonts}
\usepackage{amssymb}
\usepackage{graphicx}

\usepackage{bbm}

\def\be{\begin{equation}}
\def\ee{\end{equation}}
\def\bea{\begin{eqnarray}}
\def\eea{\end{eqnarray}}
\def\beaa{\begin{eqnarray*}}
\def\eeaa{\end{eqnarray*}}
\def\bdm{\begin{displaymath}}
\def\edm{\end{displaymath}}
\def\ba{\begin{array}}
\def\ea{\end{array}}

\newcommand{\ket}[1]{ |#1\rangle}

\newcommand{\azul}[1]{#1}
\newcommand{\rojo}[1]{#1}
\newcommand{\verdeoscuro}[1]{#1}

\newtheorem{definition}{Definition}
\newtheorem{lemma}{Lemma}
\newtheorem{proposition}{Proposition}

\begin{document}

\title{Non-Hermitian coherent states for finite-dimensional systems}
\author{Julio Guerrero$^{1,2}$}
% \address{Julio Guerrero \at Computer Science Faculty, Campus de Espinardo, University of Murcia, 30100 Murcia, Spain. \email{juguerre@um.es}\\
% Deparment of Mathematics, Experimental Sciences Faculty, Campus las Lagunillas, University of Ja\'en,  23071 Ja\'en, Spain. \email{Julio.Guerrero@ujaen.es}}

\date{}
\maketitle

% \abstract*{We introduce Gilmore-Perelomov coherent states for 
% non-unitary representations of non-compact groups, and discuss the
% main similarities and differences with respect to ordinary
% unitary Gilmore-Perelomov coherent states. The example of coherent states for
% the non-unitary
% finite dimensional representations of $SU(1,1)$ is considered and they are used 
% to describe the propagation of light in coupled PT-symmetric optical
% devices.}

\begin{center}
{\it $^1$ Deparment of Mathematics, Experimental Sciences Faculty,\\ Campus las Lagunillas, University of Ja\'en,  23071 Ja\'en, Spain.\\
JulioGuerrero@ujaen.es}\\

{\it $^2$ Computer Science Faculty, Campus de Espinardo, University of Murcia, 30100 Murcia, Spain.  }

\end{center}

\abstract{We introduce Gilmore-Perelomov coherent states for 
non-unitary representations of non-compact groups, and discuss the
main similarities and differences with respect to ordinary
unitary Gilmore-Perelomov coherent states. The example of coherent states for
the non-unitary
finite dimensional representations of $SU(1,1)$ is considered and they are used 
to describe the propagation of light in coupled PT-symmetric optical
devices.
}

\section{Introduction and physical motivation}

Non-hermitian Hamiltonians have been used for a long time as effective Hamiltonians (think, for instance, of the optical potential
in nuclear physics \cite{OpticalPotential}). With the introduction of $PT$-symmetric Hamiltonians
\cite{Bender} in Quantum Mechanics they became very popular, however only a few papers have been devoted to
coherent states (CS) for non-Hermitian systems (see \cite{Roy},
where Gazeau-Klauder CS are constructed using the definition of scalar product
in terms of the CPT norm, \cite{Trifonov,Bagarello1,Bagarello2,Bagarello3,Bagarello4,Bagarello5} where the notion of
pseudo-bosons and bi-coherent states are introduced, or \cite{Ali,Oscar}) as compared with the huge amount of papers devoted to usual 
coherent states. 

Non-hermitian systems, in particular PT-symmetric ones, have found the goose
that laid the golden eggs in optics, in particular in discrete photonic systems
(DPS), where 
the first experimental\footnote{The first theoretical proposal of $PT$-symmetry 
in optics was given in \cite{Muga}.} realization of PT-symmetry was realized
\cite{Ramy}.

Let us describe briefly DPS and later we shall study the implications on
non-hermiticity on these systems.
%

%\section{Matrix description of optical waveguide arrays}

%\subsection{Discrete photonic systems (DPS)}

 \begin{figure}[htbp]
\centering
 %trim={<left> <lower> <right> <upper>}
%\includegraphics[trim={0 12cm 0 12cm},clip,width=12cm]{Graficas/array4.png}
\includegraphics[trim={0 12cm 0 12cm},clip,width=12cm]{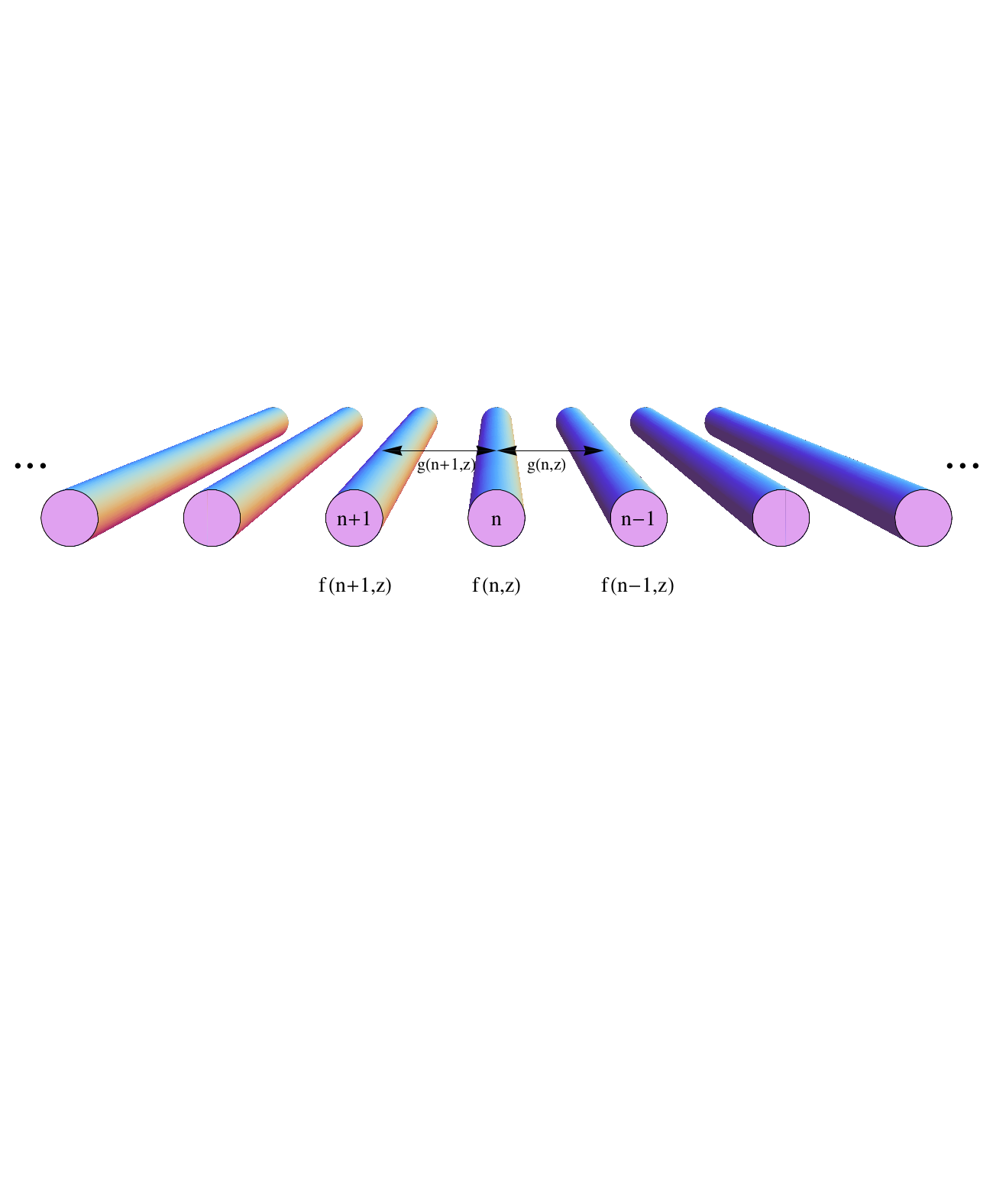}
\vspace*{-3cm}

\caption{Planar array of coupled waveguides.}
\label{DPS}
\end{figure}

Consider an (infinite) planar array of optical waveguides, like the one
of Fig. \ref{DPS}.
Using Maxwell theory of electromagnetism, and performing some
approximations (valid in a wide range of applications in DPS), we arrive to a
set of coupled first-order differential equations describing the scalar magnitude of the (either
electric o magnetic) field inside the waveguides (see for instance, \cite{CMT}):
\begin{equation}
\label{eq1}
-i\frac{d{\cal E}_n}{dz} =  f(n,z){\cal E}_n+ g(n,z){\cal
E}_{n-1}+g(n+1,z){\cal E}_{n+1}\,, \qquad {\cal E}_n(z_0)={\cal E}_n^{(0)}
%\,, n\in\mathbb{Z}
\end{equation}
where  we have supposed that the couplings between adjacent waveguides $g(n,z)$ 
(with units  inverse of length) and the propagation constant (proportional to the refractive index and also 
with units inverse of length)  $f(n,z)$
depend both on the waveguide number $n$ and the distance $z$ along the waveguide
\cite{GilmorePerelomovCS}.

Note that if the coupling $g(n,z)$ vanishes for one value of $n$ (for all $z$) the array can be semiinfinite. 
If it vanishes for two or more values of $n$ the array can be finite.

% If $g(1,z)=0$, this set of equations uncouples into two semi-infinite
% sets: $n\leq 0$ and $n\geq 1$. If further $g(N,z)=0$, for $N>0$, it uncouples
% into two seminfinite
% sets ($n\leq 0$ and $n>N$), and a finite set: $1\leq n \leq N$.

Writing eq. (\ref{eq1}) in matrix form and using Dirac's bra-ket notation we see that this set of equations can be
written in Schr\"odinger-like form (equivalence is established by simply
changing $z\rightarrow -t$):
\begin{equation}
-i\frac{d\,}{dz}\ket{{\cal E}(z)}=\hat{H}(z)\ket{{\cal E}(z)}\,, \qquad \ket{{\cal E}(z_0)}=\ket{{\cal E}^{(0)}}
\end{equation}
where
\begin{equation}
\ket{{\cal E}(z)}=\sum_{j\in {\cal I}}{\cal E}_j(z)\ket{j}\,\qquad {\cal
I}\subset \mathbb{Z}\,,
\end{equation}
%guerrero@guerrero-HP-EliteBook-820-G3:/etc/alternatives$ 

and
\bdm
\ket{j}=\left(\ba{c} \vdots\\0\\ 1 \\0 \\ \vdots \ea\right) \leftarrow
j\text{-th}\,.
\edm

The Hamiltonian $\hat{H}(z)$ reads:
\begin{equation}
\hat{H}(z)=f(\hat{n},z)+  g(\hat{n},z)\hat{V}^\dag +
\hat{V}g(\hat{n},z) \,,
\end{equation}
where $\hat{n}\ket{j}=j\ket{j}$ is the number operator, and 
$\hat{V}\ket{j}=\ket{j-1}$ and
$\hat{V}^\dag\ket{j}=\ket{j+1}$ are the step-down and step-up operators,
respectively.

% \begin{displaymath}
%    \hat{n}=\begin{pmatrix}
% \ldots&\ldots&\ldots&\ldots&\ldots& \ldots\\
% \ldots&-1&0&0&0&\ldots\\
% \ldots&0&0&0&0&\ldots\\
% \ldots& 0&0&1&0&\ldots \\\hat{U}(z_2,z_1)
% \ldots&0&0&0&2&\ldots
% \\ \ldots&\ldots&\ldots&\ldots&\ldots& \ldots \end{pmatrix}
% \end{displaymath}
% 
% 
% \begin{displaymath}
%    \hat{V}=\begin{pmatrix} 
% \ldots&\ldots&\ldots&\ldots&\ldots& \ldots\\
% \ldots&0&1&0&0&\ldots\\
% \ldots&0&0&1&0&\ldots\\ 
% \ldots&0&0&0&1&\ldots \\
% \ldots&0&0&0&0&\ldots \\
% \ldots&\ldots&\ldots&\ldots&\ldots& \ldots
% \end{pmatrix}
% \end{displaymath}
% 
% \begin{displaymath}
%    \hat{V}^\dag=\begin{pmatrix} 
% \ldots&\ldots&\ldots&\ldots&\ldots& \ldots\\
% \ldots&0&0&0&0&\ldots\\
% \ldots&1&0&0&0&\ldots\\  
% \ldots&0&1&0&0&\ldots \\ 
% \ldots&0&0&1&0&\ldots\\
% \ldots&\ldots&\ldots&\ldots&\ldots &\ldots \end{pmatrix}
% \end{displaymath}

\subsection{Symmetric DPS}

The differential equation
\be
-i\frac{d\,}{dz}\ket{{\cal E}(z)}=\hat{H}(z)\ket{{\cal E}(z)}\,,\qquad \ket{{\cal E}(z_0)}=\ket{{\cal E}^{(0)}}
\ee
can be solved by \azul{group-theoretical methods} (like Wei-Norman
factorization \cite{WeiNorman}) if
\be
\hat{H}(z)=\sum_{k=1}^N \alpha_k(z) \hat{A}_k\,,
\ee
with $\hat{A}_k$ constant matrices realizing a representation of a Lie algebra ${\cal G}$ (associated
with the Lie group $G$) with
%
%\bdm
$[ \hat{A}_i,\hat{A}_j]=\sum_{k=1}^N c_{ij}^{\,\,k}\hat{A}_k$.
%\edm
%
In this case the propagator (evolution operator in the case of Schr\"odinger
equation) can be explicitly computed:
\be
\ket{{\cal E}(z)}=U(z,z_0)\ket{{\cal E}(z_0)}=\Pi_{k=1}^N e^{i
u_k(z,z_0)\hat{A}_k}\ket{{\cal E}(z_0)}\equiv \hat{\rho}(g(z,z_0))\ket{{\cal E}(z_0)}\,,
\label{propagation}
\ee
% %
where $z_0$ is the initial value of $z$ and the functions $u_k(z,z_0)$ satisfy \rojo{non-linear first-order coupled}
differential
equations (in $z$) involving the structure
constants $c_{ij}^{\,\,k}$ and the coefficients $\alpha_k(z)$. Here  $g(z,z_0)$
represents an element of the group $G$ that, for fixed $z_0$, describes a curve in
$G$, and
$\hat{\rho}$ is a \azul{representation} of $G$ (the one obtained by
exponentiation of the representation of ${\cal G}$ defined by the matrices
$\hat{A}_k$).
%(the exponential of the representation realized by the matrices $\hat{A}_k$).

Note that the propagator satisfies the composition property 
\begin{equation}
 \hat{U}(z_2,z_1)\hat{U}(z_1,z_0)=\hat{U}(z_2,z_0)
\end{equation}
(i.e. defining a groupoid), and if the Hamiltonian $\hat{H}(z)\equiv \hat{H}$ does not depend on $z$, then 
$\hat{U}(z,z_0)=\hat{U}(z-z_0)$ and the composition property
becomes the usual homomorphysm property 
\begin{equation}
 \hat{U}(z_2)\hat{U}(z_1)=\hat{U}(z_2+z_1)\,.
\end{equation}
In this case the curve $g(z,z_0)=g(z-z_0)$
constitutes a one-parameter subgroup of $G$ generated by the Lie algebra element $\hat{H}$. 
Let's denote this subgroup by $H$.

Define as  $\ket{{\cal E}(g)}=\hat{\rho}(g)\ket{{\cal E}^{(0)}}\,,\forall g\in G$ the family of
Gilmore-Perelomov coherent states \cite{Perelomov} associated with the
group $G$, the representation
$\hat{\rho}$, and the fiducial vector $\ket{{\cal E}^{(0)}}$. Then 
$\ket{{\cal E}(z)}=\ket{{\cal E}(g(z,z_0))}\,,\forall z\in \mathbb{R}$ is a one-parameter subfamily of 
coherent states. If $\hat{H}$ does not depend on $z$, then the subfamily $\ket{{\cal E}(z)}$ is by itself
a family of coherent states associated with the subgroup $H$, the representation $\hat{\rho}$ restricted to
$H$, and the fiducial vector $\ket{{\cal E}^{(0)}}$.

An interesting fact is that, since the Hamiltonian $\hat{H}(z)$ is always an element of the 
Lie algebra ${\cal G}$, we have temporal (or rather, spatial) stability \cite{TemporallyStable}, since the 
propagation along $z$ always remains in the family $\ket{{\cal E}(g)}$ (i.e. 
$\ket{{\cal E}(z)}=\ket{{\cal E}(g)}$ for some $g\in G$). If further $\hat{H}$ does not depend on $z$ and it is
a compact operator, there will be periodic revivals, i.e. the system returns to the original state 
$\ket{{\cal E}^{(0)}}$ after multiples of some length $L$, and in general 
$\ket{{\cal E}(z+L)}=\ket{{\cal E}(z)}\,,\forall z\in\mathbb{R}$.

% Another interesting property occurs in the case where $\hat{H}(z)$ performs a ``loop'' in the Lie algebra
% ${\cal G}$, i.e., $\hat{H}(L)=\hat{H}(0)$ for some $L\in\mathbb{R}$. If $\hat{H}(z)$ varies slowly, the process

Strictly speaking, for $\ket{{\cal E}(g)}$ to be a coherent state it is
required that the representation $\hat{\rho}$ be unitary and square-integrable
(modulo a subgroup, perhaps \cite{CoherentStatesAAG}), and the fiducial
vector $\ket{{\cal E}^{(0)}}$ be admissible. However, in the examples we are
considering (and in the general discussion we provide later in Sec. \ref{nhCS}),
we shall drop the requirement of unitarity, but keeping square-integrability, and see how this still 
provides a well defined notion of coherent states.
%, and in Sec. \ref{exSU11} we shall
%analyse, for the case of finite dimensional non-unitary representations of $SU(1,1)$, 
%which are non square-integrable, in which cases (i.e. for which Hamiltonians) the resulting subfamily
%$\ket{{\cal E}(z)}$
%satisfy the square-integrability condition and therefore define true coherent
%states.

%\section{Unitarity versus non-unitarity}

We shall focus on \rojo{finite}  DPS, i.e. the index set  ${\cal I}$ will be 
\azul{finite}. If $G$ is a \azul{compact} group (like SU(2)), $U(z)$ is
\rojo{unitary} and
the total light power $ P(z)=\sum_{j\in {\cal I}}|{\cal E}_j(z)|^2$
is \verdeoscuro{conserved} along propagation.
% %
% \begin{displaymath}
%   \frac{d\,}{dz}P(z)=0
% \end{displaymath}
%
If $G$ is  a
\azul{non-compact} group (like $SO(2,1)$ or $SO(3,1)$), $U(z)$ is \rojo{not
unitary} and the
total power $P(z)$ is \verdeoscuro{not conserved}.
% \begin{displaymath}
%  \frac{d\,}{dz}P(z)\neq 0
% \end{displaymath}
%
\rojo{Non-unitarity} is caused by \rojo{non-Hermiticity} of the
Hamiltonian.%, due to complex matrix elements of the Hamiltonian.

 Non-Hermitian Hamiltonians describe DPS with
\rojo{losses} and/or
\azul{gain}, or non-symmetrical couplings (due, for instance, to torsion
of non-identical waveguides or in effective Hamiltonians for some non-linear DPS \cite{Ramy2}).

\subsection{Non-Hermitian dimer}

% \begin{figure}[htbp]
% \centering
%  %trim={<left> <lower> <right> <upper>}
% \includegraphics[width=12cm]{Graficas/PTDimer.pdf}
% %\includegraphics[height=10cm]{Graficas/array2.pdf}
% %\caption{Array of coupled waveguides}
% \end{figure}

Let us focus on a finite DPS with two waveguides, usually known as
\textit{dimer}. Consider the most general non-Hermitian coupling matrix
\be
\hat{H}_{nH}(z) = \left(\begin{matrix} \alpha_{1}(z) & \beta_{1}(z) \\ 
\beta_{2}(z) & \alpha_{2} (z) \end{matrix} \right),\qquad
\alpha_i(z)\,,\beta_j(z)\in \mathbb{C}\,.
\ee

Setting the 
%effective 
propagation constants relative to their average,
\be
\vert \mathcal{E}(z) \rangle = e^{i \int_{z_0}^{z} \alpha_{0}(t) dt } \vert E(z)
\rangle, \qquad \alpha_{0}(z) = \frac{1}{2} \left[ \alpha_{1}(z) + \alpha_{2}(z)
\right]\,,
\ee
gives the \rojo{traceless} effective non-Hermitian coupling matrix,
\be 
\hat{H}(z) = \left(\begin{matrix} \alpha(z) & \beta_{1}(z) \\  \beta_{2}(z) &
-\alpha(z) \end{matrix} \right), \qquad \alpha(z) = \frac{1}{2} \left[
\alpha_{1}(z) - \alpha_{2}(z) \right]\,,
\ee
and the differential system,
\be 
-i \partial_{z} \vert E(z) \rangle = \hat{H}(z) \vert E(z) \rangle\,.
\ee

Note that the Hamiltonian $\hat{H}(z)$ is an element of the  $sl(2,\mathbb{C})\approx
so(3,1)$ Lie algebra,
where $sl(2,\mathbb{C})=so(3)\oplus i\, so(3)$, with basis
$\{\hat{J}_x,\hat{J}_y,\hat{J}_z,i\hat{J}_x,i \hat{J}_y,i \hat{J}_z\}$,
$\hat{J}_k$ being the standard \azul{angular momentum} operators (Pauli
matrices in the case of the dimer). More precisely, it can be written as:
\begin{eqnarray}
 \hat{H}(z) &=& \Re(\beta_1(z)+\beta_2(z))\hat{J}_x +\Im(\beta_1(z)+\beta_2(z)(i \hat{J}_x)\nonumber \\
 & & + \Im(\beta_2(z)-\beta_1(z))\hat{J}_y+\Re(\beta_1(z)-\beta_2(z))(i \hat{J}_y)\\
 & & + 2\Re(\alpha(z)) \hat{J}_z + 2\Im(\alpha(z)) (i \hat{J}_z) \nonumber 
\end{eqnarray}

Designing appropriately the parameters in the Hamiltonian  some of the coefficients in the right hand side of previous 
equation can vanish and thus $\hat H(z)$ will be an element of the Lie subalgebra $su(1,1)\approx
so(2,1)\subset so(3,1)$, whose  generators are denoted by
$\{\hat{K}_{x},\hat{K}_{y},\hat{K}_{z}\}$. For this purpose, there are various
possibilities:
\be
% 
% Varios conjugate $so(2,1)$ subalgebras can be found in $sl(2,\mathbb{C})$:
% \vskip 0.3cm
\{\hat{K}_{x}, \hat{K}_{y},\hat{K}_{z} \} \equiv \{i\hat{J}_{x},
i\hat{J}_{y},\hat{J}_{z}\}, ~ \{ i\hat{J}_{y}, i\hat{J}_{z},\hat{J}_{x}\},
~\{i\hat{J}_{z}, i\hat{J}_{x},\hat{J}_{y}\}  \ldots\,.
\ee

\verdeoscuro{Experimentally}, it is easier to realize a Hamiltonian that can be written as 
a combination of 
$\{\hat{K}_{x},\hat{K}_{y},\hat{K}_{z}\} \equiv \{i \hat{J}_{y},
i \hat{J}_{z},\hat{J}_{x}\}$, imposing $\Re(\alpha(z))=0$, $\Im(\beta_1(z)+\beta_2(z))=0$ and 
$\Im(\beta_2(z)-\beta_1(z))=0$ (i.e. $\beta_1(z)$, $\beta_2(z)$ real and $\alpha(z)$ pure imaginary). If we further impose $\Re(\beta_1(z)-\beta_2(z))=0$, i.e. 
$\beta_1(z)=\beta_2(z)$ and real (i.e. real and symmetrical couplings), this system corresponds to a \rojo{PT-symmetric} DPS
with balanced \azul{gain}/\rojo{loss} \cite{OptLett,Symmetry}. In the next subsection a realization
of this case is discussed in detail.

\subsection[PT-Symmetric waveguide arrays]{Finite DPS with
$SO(2,1)$ symmetry: PT-symmetric DPS}

The Hamiltonian for a finite DPS with underlying
$SO(2,1)\approx SU(1,1)$ symmetry can be taken as
\begin{equation}
\hat{H}(z) = i \gamma(z) \hat{J}_{z} + \lambda(z) \hat{J}_x \equiv \gamma(z)
\hat{K}_{y} + \lambda(z) \hat{K}_z\,.
\label{Hamiltonian}
\end{equation}

If we take a unitary representation of $so(3)\approx su(2)$ with spin $j\in\frac{\mathbb{Z}}{2}$, we obtain 
a non-unitary representation of $SO(2,1)\approx SU(1,1)$ with ``spin'' 
$j\in\frac{\mathbb{Z}}{2}$. Since this representation is finite dimensional with dimension $2j+1$,
our system will have $2j+1$ waveguides.

% In this case  there are $2j+1$ waveguides for the non-unitary representation with
% ``spin'' 
% $j\in\frac{\mathbb{Z}}{2}$ of $SO(2,1)\approx SU(1,1)$.

Under the \rojo{$P$ transformation} ($n\rightarrow 2j -n$) and \azul{``$T$''
transformation} ($i\rightarrow -i$, $z\rightarrow -z$) the Hamiltonian changes
to:
\be
\hat{H}(z) \rightarrow \hat{H}(z)^{PT}=- i \gamma(-z) \left(-\hat{J}_{z}\right)
+ \lambda(-z) \hat{J}_x  = \hat{H}(z)\,,
\ee
provided $\gamma(z)$ and $\lambda(z)$ are even. Therefore the system is
\rojo{$PT$-symmetric}.

However, for some values of the parameters, the \azul{$PT$-symmetry}  can be
\rojo{spontaneously broken}, in the sense that
 \azul{wavefunctions are not PT-symmetric}.

\begin{figure}
\includegraphics[width=\linewidth]{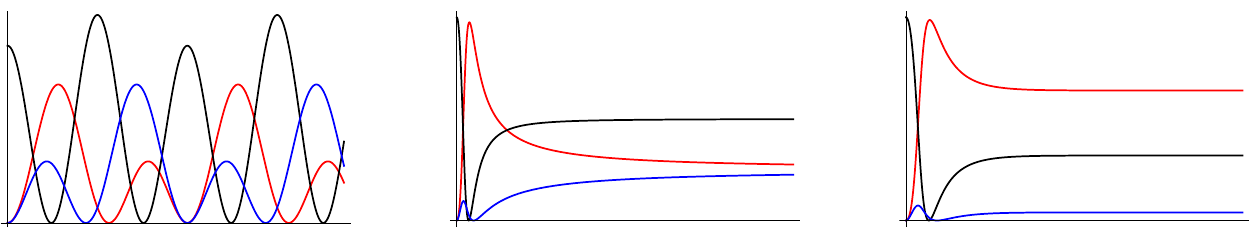}     
\caption{Plots of light intensity for the PT-symmetric DPS with 3 waveguides for
the cases $\gamma<\lambda$  (left), $\gamma=\lambda$ (center) 
and  $\gamma>\lambda$ (right). Light is impinged in the central waveguide
(in black), while the waveguide with gain is shown in red and the one with
loss in blue. In the last two cases light intensity has been renormalized to
show the asymptotic behavior.}
\label{PTtrimer}
\end{figure}

For the case $\gamma(z)=\gamma$ and $\lambda(z)=\lambda$, simple analytical
expressions can be obtained. Three cases
have to be considered, depending on whether $\Omega=\sqrt{\lambda^2-\gamma^2}$
is \azul{positive}, \verdeoscuro{zero} or \rojo{pure imaginary}, corresponding
to three qualitatively different propagations:
\azul{periodic} (but with total power not conserved along propagation, unbroken PT-symmetry),
growing with a
\verdeoscuro{power law}, or
growing with an \rojo{exponential law}, respectively (broken PT-symmetry). See
Fig. \ref{PTtrimer} showing the light intensities for a PT-symmetric DPS with
three waveguides.

Guided by the physical motivation of non-hermitian DPS, we shall try to provide
in the rest of the paper a mathematical framework that describes the propagation
of light in these devices as non-hermitian (or non-unitary)
Gilmore-Perelomov coherent states.

\section{Non-Hermitian Coherent States}
\label{nhCS}

% The previous considerations can be mathematically formulated in the context of
% coherent states for \rojo{non-unitary} \azul{finite-dimensional} representations
% of a \verdeoscuro{non-compact} group.
% 

As we have seen in the physical example, in some situations one has to handle systems possessing a symmetry
group that it is realized through a non-unitary representation. To
mathematically describe this situation, one has 
to resort (in the infinite dimensional case) to a more general setting than Hilbert spaces, and this is the
notion of Banach space. However, to simulate some of the properties of unitary representations on Hilbert spaces, a ``dual'' representation is needed.
Let us introduce the notion of contragredient representation $\hat{\rho}^*$ associated to a non-unitary representation $\hat{\rho}$, following \cite{Zimmerman}. 

Let $\hat{\rho}$ be a representation  of a locally compact Lie group $G$
with Lie algebra ${\cal G}$ in a reflexive\footnote{We shall restrict to reflexive Banach spaces since
in this case the formulation is simpler and more similar to the Hilbert space case, see \cite{Zimmerman}.} Banach  space ${\cal V}$:
\be
\hat{\rho}:\,G\rightarrow GL({\cal V})\,.
\ee

Define the contragredient  representation $\hat{\rho}^*$ on the dual space ${\cal V^*}$, $\hat{\rho}^*:\,G\rightarrow GL({\cal V^*})$ as $\hat{\rho}^*(g)=\hat{\rho}(g^{-1})^\dag$.

Note that, since  ${\cal V}$ is assumed to be reflexive, we have that if $\hat{\rho}$ is bounded and continuous,
then $\hat{\rho}^*$ is continuous, and that $\hat{\rho}$ is irreducible if and only if $\hat{\rho}^*$ is irreducible.

By definition, the pair $(\hat{\rho},\hat{\rho}^*)$ verifies
\be
\langle \hat{\rho}^*(g)\tilde{\Psi},\hat{\rho}(g)\Phi\rangle = \langle
\tilde{\Psi},\Phi\rangle\,\qquad \forall \tilde{\Psi}\in {\cal V^*},\,\forall
\Phi\in {\cal V}\,,
\ee
where $\langle\tilde{\Psi},\Phi\rangle=\tilde{\Psi}(\Phi)$ denotes the usual pairing between ${\cal V}$ and ${\cal V^*}$.

Note that if  $\hat{\rho}$ is unitary (and ${\cal V}$ is a Hilbert space, where $\langle\cdot,\cdot\rangle$ is
the scalar product, related to the pairing by Riesz representation theorem) then $\hat{\rho}^*=\hat{\rho}$. In this sense, the pair $(\hat{\rho},\hat{\rho}^*)$ generalizes the unitary case to the non-unitary case, in the more
general setting of Banach spaces.

The notion of square integrability of a representation should be adapted to this more general setting as follows:
\begin{definition} (\cite{Zimmerman})
 A representation $\hat{\rho}$ on a reflexive Banach space ${\cal V}$ is square integrable if there exist a
 non-zero vector $\Phi\in V$ (called admissible) such that
 \be
 \int_G |\langle \tilde{\Psi},\hat{\rho}(g)\Phi\rangle|^2 d\mu(g)
<\infty\,,\qquad\forall \tilde{\Psi}\in{\cal V}^* \,,
 \ee
 where $d\mu(g)$ is the (left) Haar measure on $G$.
\end{definition}

An analogous definition can be given for the contragredient representation:
\begin{definition}  (\cite{Zimmerman})
 The contragredient representation $\hat{\rho}^*$ on ${\cal V}^*$ is ${}^*$-square integrable if there exist a
 non-zero vector $\tilde{\Psi}\in V^*$ (called ${}^*$-admissible) such that
 \be
 \int_G |\langle \hat{\rho}^*(g)\tilde{\Psi},\Phi\rangle|^2 d\mu(g)
<\infty\,,\qquad\forall \Phi\in{\cal V}\,,
 \ee
 where $d\mu(g)$ is the (left) Haar measure on $G$.
\end{definition}

The set of admissible (resp. ${}^*$-admissible) vectors is invariant under $\hat{\rho}$ (resp. $\hat{\rho}^*$).

From now on we shall say, to simplify the notation, that the representation $\hat{\rho}^*$ is
square integrable if it is ${}^*$-square integrable and that the non-zero vector $\tilde{\Psi}\in {\cal V}^*$ is admissible if it is
${}^*$-admissible.

\subsection{Non-hermitian Gilmore-Perelomov Coherent States}

Let us generalize the notion of Gilmore-Perelomov coherent states
\cite{Perelomov,CoherentStatesAAG} to this wider setting.

\begin{definition}
 Given a square integrable representation $\hat{\rho}$ on a reflexive Banach space ${\cal V}$ and an
admissible vector $\Phi\in {\cal V}$,  a family of non-hermitian Gilmore-Perelomov Coherent States
is given by:
\be
 \Phi_g =\hat{\rho}(g)\Phi\,,\quad \forall g\in G\,.
\ee
\end{definition}

The analogous ``dual'' definition is:

\begin{definition}
 Given the square integrable representation $\hat{\rho}^*$ on ${\cal V}^*$ and an
admissible vector $\tilde{\Psi}\in{\cal V}^*$,  the dual family of non-hermitian Gilmore-Perelomov Coherent States
is given by:
\be
 \tilde{\Psi}_g =\hat{\rho}^*(g)\tilde{\Psi}\,,\quad \forall g\in G\,.
\ee
\end{definition}

The idea of defining two families of coherent states can be seen as a
generalization (in the sense of Gilmore-Perelomov coherent states) of
bi-coherent states for the standard bosonic operators \cite{Trifonov,Bagarello1,Bagarello2,Bagarello3,Bagarello4,Bagarello5}.

The notion of square integrability (and its dual) is required to make sense of the following concepts (to avoid cluttering of notation we shall omit the
dependence on the representation and the admissible vector on $\hat{T}$ and $\hat{\widetilde{T}}$).

\begin{definition}  (\cite{Zimmerman}) Let $\hat{\rho}$ be a square integrable representation of the locally compact group $G$ on the reflexive Banach space
${\cal V}$, and $\Phi\in {\cal V}$ an admissible vector for  $\hat{\rho}$. Then the \textit{sampling} (or \textit{analysis}) operator is given by
 
\begin{eqnarray}
\hat{T}:\, {\cal V}^* &\rightarrow& L^2(G) \nonumber\\
   \tilde{\Psi} &\mapsto&    \hat{T}(\tilde{\Psi})(g)=\overline{\langle
\tilde{\Psi},\Phi_g\rangle}\,.
\end{eqnarray}

The adjoint defines the \textit{synthesis} operator:
\begin{eqnarray}
\hat{T}^\dag:\,L^2(G)  &\rightarrow&  {\cal V}  \nonumber\\
   \gamma &\mapsto&    \hat{T}^\dag(\gamma)=\int_G \gamma(g)\Phi_g d\mu(g)\,.
\end{eqnarray}
\end{definition}

There is an analogous definition for the dual case:

\begin{definition}  (\cite{Zimmerman}) Let $\hat{\rho}^*$ be a square integrable representation of the locally compact group $G$ 
on ${\cal V}^*$, and $\tilde{\Psi}\in {\cal V}^*$ an admissible vector for  $\hat{\rho}^*$. Then the dual \textit{sampling} (or \textit{analysis}) operator is given by
 
\begin{eqnarray}
\hat{\widetilde{T}}:\, {\cal V} &\rightarrow& L^2(G) \nonumber\\
   \Phi &\mapsto&    \hat{\widetilde{T}}(\Phi)(g)=\langle
\tilde{\Psi}_g,\Phi\rangle\,.
\end{eqnarray}

The adjoint defines the synthesis operator:
\begin{eqnarray}
\hat{\widetilde{T}}^\dag:\,L^2(G)  &\rightarrow&  {\cal V}^*  \nonumber\\
   \gamma &\mapsto&    \hat{\widetilde{T}}^\dag(\gamma)=\int_G \gamma(g)\tilde{\Psi}_g
d\mu(g)\,.
\end{eqnarray}

\end{definition}

%%%%%%%%%%%%%%%%%%%%%%%%%%%%%%%%%%%%%%%%%%%%%%%%%%%%%%%%%%%%5

 Let us introduce the Resolution operator $\hat{A}=\hat{T}^\dag\hat{\tilde{T}}$:
 \begin{eqnarray}
  \hat{A}:\,{\cal V} &\rightarrow&  {\cal V} \nonumber\\
  \phi &\mapsto&\hat{T}^\dag\hat{\tilde{T}}\phi= \int_G \langle
\tilde{\Psi}_g,\phi\rangle \Phi_g d\mu(g)\,,
 \end{eqnarray}
%
% 
% 
 %\be
 %\hat{A}= \hat{T}^\dag\hat{\tilde{T}}=\int_G d\mu(g) \hat{\rho}(g) |\Phi\rangle
%\langle \Psi|\hat{\rho}(g)^{-1}\,,\,
 %\end{displaymath}
 %\begin{displaymath}
 % 
and its dual $\hat{\tilde{A}}=\hat{A}^\dag= 
\hat{\tilde{T}}^\dag\hat{T}$.
%= \int_G d\mu(g) \hat{\rho}^*(g) |\Phi\rangle \langle \Psi|\hat{\rho}(g)^\dag
 %\ee

A non-unitary version of Schur's lemma can be stated. For this purpose, denote by $L({\cal V})$ the Banach algebra of bounded linear operators on the Banach space ${\cal V}$.
We need the following results, that can be
obtained by a slight modification of those in \cite{Aupetit} (Corollary 3.2.9 and Theorem 4.2.2).

% \begin{lemma}
%  If $A$ is a Banach algebra in which every nonzero element is invertible then it is 
%  isometrically isomorphic to $\mathbb{C}$.
% \end{lemma}

\begin{lemma}
Let $\hat{\rho}$ be a bounded and irreducible representation of the Lie group $G$ on the 
Banach space ${\cal V}$. If we define  
 $C=\{\hat{B}\in L({\cal V}):\,\hat{\rho}(g)\hat{B}=\hat{B}\hat{\rho}(g)\,,\,\forall g\in G\}$,
 then $C$ is isomorphic to $\mathbb{C}$.
\end{lemma}

% \noindent \textit{Proof:}
% From the definition of $C$ it is immediate to prove that $Ker(\hat{B})$ and $Im(\hat{B})$ are 
% invariant subspaces for $\hat{\rho}$. Since $\hat{\rho}$ is irreducible, either 
% $Ker(\hat{B})=\{0\}$ and 
%  $Im(\hat{B})={\cal V}$ or $Ker(\hat{B})={\cal V}$ and 
%  $Im(\hat{B})=\{0\}$. In the first case $\hat{B}$ is invertible and in the second case 
%  $\hat{B}$ is identically zero (and therefore proportional to the identity operator with
%  proportionality constant $\lambda=0$). Suppose that $\hat{B}$ is invertible, by the Open Mapping Theorem 
%  $\hat{B}^{-1}$ is bounded, and verifies $\hat{\rho}(g)\hat{B}^{-1}=\hat{B}^{-1}\hat{\rho}(g)\,,\,\forall g\in G$,
%  and thus $\hat{B}^{-1}\in C$. Thus $C$ is a closed Banach subalgebra of the Banach algebra
%  $L({\cal V})$ verifying that for every non-zero element $\hat{B}\in C$, $\hat{B}^{-1}\in C$. By
%  Corollary 3.2.9 of \cite{Aupetit} $C$ is isomorphic to $\mathbb{C}$.
%  

With this Lemma  we can prove:

% We shall give a simpler version
% of it, easier to proof, which is however enough for the example we shall consider in Sec. 
% \ref{exSU11}.

\begin{proposition} 
\label{Prop1} Let ${\cal V}$ be a reflexive Banach space. If $\hat{\rho}$ is continuous and irreducible, the resolution operators
$\hat{A}$ and $\hat{\tilde{A}}$, if non-zero and bounded, are proportional to the identity operator on
${\cal V}$ and ${\cal V}^*$, respectively.
\end{proposition}

\noindent \textit{Proof:} Using the definitions of $\hat{A}$ and
$\hat{\tilde{A}}$ and the invariance of the left Haar measure on $G$, it is easy
to check that
 \begin{equation}
 %\hat{\tilde{A}}=\hat{A}^\dag\,,\quad
\hat{\rho}(g)\hat{A}=\hat{A}\hat{\rho}(g)\,,\quad\hat{\rho}^*
(g)\hat{\tilde{A}}=\hat{\tilde{A}}\hat{\rho}^*(g)\,,\quad\forall g\in G\,.
\label{interwinning}
 \end{equation}
 
 Thus  $\hat{A}\in C$ and $\hat{\tilde{A}}\in C^*$ (defined analogously for $\hat{\rho}^*$). 
 Since $\hat{\rho}$ is irreducible also $\hat{\rho}^*$ is and by the previous Lemma 
 both $C$ and 
 $C^*$ are isomorphic to $\mathbb{C}$, and therefore $\hat{A}$ and $\hat{\tilde{A}}$ are proportional to the 
 identity operators. 
%
%  From this it is immediate to prove that $Ker(\hat{A})$ and $Im(\hat{A})$ are invariant subspaces
%  for $\hat{\rho}$.
%  If the representation $\hat{\rho}$ is irreducible, either $Ker(\hat{A})=\{0\}$ and 
%  $Im(\hat{A})={\cal V}$ or $Ker(\hat{A})={\cal V}$ and 
%  $Im(\hat{A})=\{0\}$. In the first case $\hat{A}$ is invertible and in the second case 
%  $\hat{A}$ is identically zero (and therefore proportional to the identity operator with
%  proportionality constant $\lambda=0$). Suppose that $\hat{A}$ is invertible, then it is 
%  immediate that $\hat{A}^{-1}$ verifies $\hat{\rho}(g)\hat{A}^{-1}=\hat{A}^{-1}\hat{\rho}(g)\,,\,\forall g\in G$.
%  
%  
%  
%  by  Lemma 0.5.2 of \cite{Wallach} we have that
% $\hat{A}=\lambda I_{\cal V}$. Assuming that ${\cal V}$ is reflexive as before we have that 
% $\hat{\rho}^*$ is also irreducible and therefore we also have that 
% $\hat{\tilde{A}}=\lambda^* I_{{\cal V}^*}$.
$\blacksquare$

It should be stressed that the requirement that $\hat{A}$ (and thus $\hat{\tilde{A}}$)
be bounded is crucial for the validity of the previous results (otherwise $\hat{A}\notin C$ and it cannot
be granted that $\hat{A}^{-1}$ is bounded). Also, the 
requirement of irreducibility is necessary to have that  $\hat{A}$ is bijective (and not just that $\hat{A}$ is injective with 
$\hat{A}^{-1}$ having dense domain in ${\cal V})$.

If $(\hat{\rho},\hat{\rho}^*)$ are square-integrable (with
their respective admissible vectors), then we have that $\hat{A}$ and $\hat{\tilde{A}}$ are
bounded (see \cite{Zimmerman}), and therefore Proposition \ref{Prop1} applies.

Similarly to the unitary case, a reproducing (or overlapping) kernel can be defined.

\begin{definition}
\label{overlapkernel}
 Let ${\cal V}$ be a reflexive Banach space and  $\hat{\rho}$ a continuous representation
 of a Lie group $G$ on ${\cal V}$, with $\hat{\rho}^*$ its contragredient representation.  Define the \rojo{Overlapping Kernel} as:
\be
\kappa(g',g)=\langle \tilde{\Psi}_{g'},\Phi_g\rangle =\langle
\hat{\rho}^*(g')\Psi,\hat{\rho}(g)\Phi\rangle \,.
\ee

\end{definition}

\begin{proposition} The overlapping kernel $\kappa(g',g)$ depends only on
$g'^{-1}g$.
\label{convkernel}
\end{proposition}
  
\noindent \textit{Proof:}
\be
\kappa(g',g) =\langle
\hat{\rho}^*(g')\Psi,\hat{\rho}(g)\Phi\rangle  = \langle
\Psi,\hat{\rho}(g')^{-1}\hat{\rho}(g)\Phi\rangle = \kappa(e, g'^{-1}g)\equiv
\kappa(g'^{-1}g)\,,
\ee
where $e$ is the identity element in $G$. $\blacksquare$

Proposition \ref{convkernel} means that $\kappa(g',g)$ is the kernel of a convolution operator on $L^2(G)$ when $\kappa(g',g)$ is square integrable.

The overlapping kernel, in the unitary and square-integrable case, endorses the Hilbert subspace $Im(\hat{T})\subset L^2(G)$
with a reproducing kernel Hilbert space structure, with kernel $\kappa(g',g)$, which is square-integrable, hermitian ($\kappa(g,g')=\overline{\kappa(g',g)}$)
and positive definite.
In the non-unitary case, if $(\hat{\rho},\hat{\rho}^*)$ are square-integrable, we still have that
the overlapping kernel is square-integrable, but since ${\cal V}^*\neq {\cal V}$, we need two
admissible functions $(\Phi,\tilde{\Psi})$ and therefore $\kappa(g',g)$ and $\kappa(g,g')$ in general are not 
related to each other and does not make sense to speak about hermiticity and positive definiteness.
% In certain cases, if there is an isomorphism $\hat{\nu}:{\cal V}\rightarrow {\cal V}^*$ such that
% $\tilde{\Psi}=\hat{\nu}\Phi$, then
Therefore, $\kappa(g',g)$ does not endorses $Im(\hat{T})$ with a reproducing kernel Hilbert space structure.

\subsection{The non-square integrable case}
 
If the representations $(\hat{\rho},\hat{\rho}^*)$ are not square integrable, previous expressions are purely formal in the sense that the norm of
$\Phi_g$ and $\widetilde{\Psi}_g$ can be \rojo{unbounded} on $G$
(implying $\hat{A}$ unbounded  and $\kappa(g',g)$ non-integrable) due to the
non-unitarity of $\hat{\rho}$ and non-compactness of $G$.

This means that, although it is possible to define a family of coherent states, it is useless in the
sense that we cannot use it as a ``basis'', and in particular it is not possible to reconstruct a function
$\phi$ from its coefficients $\langle \tilde{\Psi}_g,\phi\rangle$ in this ``basis''. This process requires
the inverse of the resolution operator $\hat{A}$ and this operator is not bounded (and even can be
``infinite'' in the sense that the integral on $G$ defining it is divergent).

Despite of these problems related to the lack of square-integrability, non-hermitian Coherent
States can be \azul{physically meaningful} in some situations, as shown in previous sections.

The reason can be understood as follows. Suppose $G$ is a non-compact Lie group and $\hat{\rho}$ a non-unitary representation, with 
$\hat{\rho}^*$ its contragredient representation. Suppose they are not square integrable and consider
the maximal compact subgroup $K$ of $G$. Since $K$ is compact, the restriction of the 
representations $(\hat{\rho}|_K,\hat{\rho}^*|_K)$ are both equivalent to a unitary representation
$\hat{\sigma}$ of $K$. Redefine $(\hat{\rho},\hat{\rho}^*)$ such that restricted to $K$ coincide with 
$\hat{\sigma}$. Since $K$ is compact $\hat{\sigma}$ is square-integrable, thus consider $(\Phi^K,\tilde{\Psi}^K)$ admissible
vectors in ${\cal V}$ and ${\cal V}^*$, respectively. The resolution operator $\hat{A}_K$ is
\begin{equation}
\label{AK}
 \hat{A}_{K}\phi=\int_{K} \langle \tilde{\Psi}^K_k,\phi\rangle \Phi^K_k d\mu_K(k)\,,
\end{equation}
where $d\mu_K(k)$ is a left-invariant Haar measure on $K$ and 
$(\Phi^K_k,\tilde{\Psi}^K_k)=(\hat{\sigma}(k)\Phi^K,\hat{\sigma}(k)\tilde{\Psi}^K)$.
$A_K$ is bounded and satisfy the interwinning
property, Eq. (\ref{interwinning}), with respect to $\hat{\sigma}$. In general $\hat{\sigma}$ is not irreducible (even if $\hat{\rho}$ is) and therefore we cannot apply
Prop. \ref{Prop1} to conclude that $\hat{A}_{K}$ is proportional to the identity operator
on ${\cal V}$ (the same applies to $\hat{\tilde{A}}_K$ on ${\cal V}^*$). 

%The overlapping kernel restricted to $K$, $\kappa_K(k',k)$, is 

Consider now $X=G/K$ and define the \emph{elliptic} subgroups $K_{[v]}=\{k_v=v k v^{-1}\,,k\in K\}$, where $v\in G$ is 
a representative for a class $[v]\in X$. Define $(\hat{\rho}_{[v]},\hat{\rho}^*_{[v]})$ as the restriction
to $K_{[v]}$ of $(\hat{\rho},\hat{\rho}^*)$. Then we have that 
%
%\begin{proposition}
 the representations $(\hat{\rho}_{[v]},\hat{\rho}^*_{[v]})$ are square integrable and the resolution operator $\hat{A}_{K_{[v]}}$ is given
 by:
 %\begin{equation}
$ \hat{A}_{K_{[v]}}\phi= \hat{\rho}(v) \hat{A}_K \hat{\rho}(v^{-1})$\,.

As before, 
in general $(\hat{\rho}_{[v]},\hat{\rho}^*_{[v]})$ are not irreducible (even if $\hat{\rho}$ is) and therefore we cannot apply
Prop. \ref{Prop1} to conclude that $\hat{A}_{[v]}$ is proportional to the identity operator
on ${\cal V}$. 

These considerations can even be extended to the general case. Let  $\hat{\rho}$ be a non-unitary
representation of the non-compact group $G$ on the Banach space ${\cal V}$. If $\hat{\rho}$
is square integrable, then ${\cal V}$ is topologically isomorphic to a Hilbert space ${\cal H}$ 
and $(\hat{\rho},\hat{\rho}^*)$ are equivalent to a unitary representation on ${\cal H}$ (see \cite{Zimmerman}, Theorem 3.13).
This result can be seen as a group-theoretical version of the concept of pseudo-hermiticity 
(see \cite{Mostafazadeh}).

In the next section, motivated by the $PT$-symmetric dimer, we consider the example of the non-unitary finite-dimensional representations of 
$SU(1,1)$, and discuss various ways of tackling the lack of square-integrability of the representations.
 
%  \rojo{Square integrability} of the representation $\hat{\rho}$ is required
%(and
% admissibility conditions for the vectors $|\Psi\rangle$ and $|\Phi\rangle$),
%or
% a suitable restriction in the integration to a quotient space $G/H$ or to a
% subset $C\subset G$, but in this case the resulting resolution operator
%  need not verify $\hat{A}=\lambda I_{\cal H}$.
% 
% 
% 
% Define the \rojo{Overlapping Kernel}:
% %
% \begin{eqnarray*}
% K(g',g)&=&\langle \tilde{\Phi}_{g'}|\Psi_g\rangle =\langle
%\Phi|\hat{\rho}^*(g')^\dag\hat{\rho}(g)\Psi\rangle \\
% &=& \langle \Phi|\hat{\rho}(g')^{-1}\hat{\rho}(g)\Psi\rangle = K(g'^{-1}g)
% \end{eqnarray*}
% 
% 
%  
%  Under suitable conditions $K(g',g)$ is a reproducing kernel defining 
%  a reproducing kernel Hilbert space.
% %The reproducing Kernel is positive definite if $|\Phi\rangle = |\Psi\rangle$.
% 

%\section{Example: SU(1,1) Non-hermitian coherent states}
\section{Example: SU(1,1) Non-hermitian coherent states}
\label{exSU11}

Consider the realization (see \cite{Perelomov}) of $su(1,1)\approx so(2,1)$ 
$\{\hat{K}_x\equiv i \hat{J}_x,\hat{K}_y\equiv i \hat{J}_y, \hat{K}_z=\hat{J}_z\}$.
Note that this is a different realization to that of the $PT$-symmetric dimer, we shall use this one in this
example since the compact generator of $SU(1,1)$ is  $\hat{K}_z=\hat{J}_z$, the usual convention in
mathematics. All results discussed here apply to the case of the $PT$-symmetric dimer by changing
\begin{equation}
\label{change}
\hat{J}_z\rightarrow \hat{J}_x\,\,\qquad \hat{J}_y\rightarrow \hat{J}_z\,, \qquad 
\hat{J}_x\rightarrow \hat{J}_y\,.
\end{equation}

%\vskip 0.3cm

Thus, if we take   the $2j+1$-dimensional unitary irreducible
representation of $SU(2)$, we get
a $2j+1$-dimensional \rojo{non-unitary } irreducible representation
of $SU(1,1)$.

%\vskip 0.3cm

We shall use the following parametrization of the group elements (for the case $j=1/2$, leading to the fundamental o defining representation,
 for other values of $j$ provides the different non-unitary representations of the group):
\begin{equation}
\hat{\rho}(\zeta,\zeta^*,\beta)=e^{\xi \hat{K}_+-\xi^*\hat{K}_-}e^{i\beta \hat{K}_z}\,,
\quad\xi\in\mathbb{C}\,,
\quad  \zeta=\frac{\xi}{|\xi|} \tanh |\xi|\,,
\label{repSU11}
\end{equation}
where $ \zeta\in\mathbb{D}$ (the unit disk in the complex plane), $\beta \in [0,4\pi)$ and
$\hat{K}_\pm=-\hat{K}_y\pm i \hat{K_x}$. Note that $\hat{K}_\pm$ are not the adjoint of each other, but instead 
$\hat{K}_+^\dag=-\hat{K}_-$. For this reason $e^{\xi \hat{K}_+-\xi^*\hat{K}_-}$ is not a unitary operator, in fact it is
hermitian positive-definite. Thus eq. (\ref{repSU11}) provides explicitly the polar decomposition of the representation, and
shows that it is non-unitary.

For instance, the $j=1/2$ case is:
\be
\hat{\rho}_{1/2}(\zeta,\zeta^*,\beta)=\frac{1}{\sqrt{1-|\zeta|^2}}
\left(
\begin{array}{cc}
 e^{i \beta/2 } & e^{i \beta/2} \zeta  \\
 e^{-i \beta/2 }\zeta^*& e^{-i \beta/2 }
\end{array}
\right)\,.
\ee

The contragredient representation for this case is:
\be
\hat{\rho}_{1/2}^*(\zeta,\zeta^*,\beta)=\frac{1}{\sqrt{1-|\zeta|^2}}
\left(
\begin{array}{cc}
 e^{i \beta/2 } & -e^{i \beta/2} \zeta  \\
 -e^{-i \beta/2 }\zeta^*& e^{-i \beta/2 }
\end{array}
\right)\,.
\ee

It is important to mention that, for general $j$, matrix elements of both representations diverge as $(1-|\zeta|)^{-j}$ when 
$|\zeta|\rightarrow 1$. This is also the behavior of the largest eigenvalue.

Let us choose as fiducial vectors (other choices lead to similar results in this case) 
$\Phi=\tilde{\Psi}=\left( \begin{array}{c} 1 \\ 0 \end{array}  
\right)$.
%\vskip 0.3cm
Then the coherent states are:
\begin{eqnarray*}
\Phi(\zeta,\beta)_{1/2}& =&\frac{1}{\sqrt{1-|\zeta|^2}}\left( \begin{array}{c}e^{i\beta/2} \\ 
 e^{-i\beta/2}\zeta^*\end{array}   \right) \\
 \widetilde{\Psi}(\zeta,\beta)_{1/2}& =&\frac{1}{\sqrt{1-|\zeta|^2}}\left( \begin{array}{c}e^{i\beta/2} \\ 
 -e^{-i\beta/2}\zeta^*\end{array}   \right) \,.
\end{eqnarray*} 

The resolution operator is:
\be
\hat{A}_{1/2}=\int_{\mathbb{D}\times [0,4\pi)} \frac{d\zeta d\zeta^*}{(1-|\zeta|^2)^2}\frac{d\beta}{4\pi} \frac{1}{1-|\zeta|^2}
\left(
\begin{array}{cc}
 1 & e^{i \beta} \zeta  \\
 e^{-i \beta }\zeta^*& -|\zeta|^2
\end{array}
\right)\,.
\ee
where $\frac{d\zeta d\zeta^*}{(1-|\zeta|^2)^2}\frac{d\beta}{4\pi}$ is the left-invariant Haar measure on the group. The dual version is simply 
the adjoint. $\hat{A}_{1/2}$
 is divergent since the representation is not square integrable. The reason is
that the behavior for $|\zeta|\rightarrow 1$ 
of the matrix elements of $\hat{A}_j$,  for arbitrary $j$,  is
 $(1-|\zeta|)^{-2j}$, and together with the integration measure gives $(1-|\zeta|)^{-2j-2}$ which is divergent when integrated 
on $\mathbb{D}$.

This should be compared with the unitary representations of the discrete series of $SU(1,1)$, which are infinite-dimensional, and the
behavior for $|\zeta|\rightarrow 1$ of the matrix elements of the corresponding
resolution operator $\hat{A}_k$ is $(1-|\zeta|)^{2k}$, 
where $k$ is the Bargmann index characterizing the representation.
Together with the integration measure gives $(1-|\zeta|)^{2k-2}$ and this is convergent when integrated 
on $\mathbb{D}$ if $k>1/2$. Therefore, the representations of the discrete series are unitary and square-integrable for $k>1/2$.

Note that all expressions can be derived from those of the discrete series simply 
performing the change $k\rightarrow -j$, with $j>0$ and half-integer.

The overlapping kernel is given by:
\be
\kappa_{1/2}((\zeta',\beta'),(\zeta,\beta))=\frac{1}{\sqrt{(1-|\zeta|^2)(1-|\zeta'|^2)}}
(e^{i \frac{\beta-\beta'}{2}}-e^{-i \frac{\beta-\beta'}{2}}\zeta^*\zeta')\,.
\ee

In this case the overlapping kernel is hermitian and positive definite (since ${\cal V}^*$ is isomorphic to ${\cal V}$ 
and $\Phi=\tilde{\Psi}$), but it is not bounded on $\mathbb{D}$ and therefore it is not square-integrable. In fact, as commented previously, it
coincides with the
overlapping kernel for the coherent states associated with the representation of
the discrete series with Bargmann index \rojo{$k=-1/2$}.

%\vskip 0.3cm

To overcome these problems there are many alternatives, as suggested in Sec. \ref{nhCS}. We shall discuss some of them in the following subsections.

\subsection{Restriction to the maximal compact subgroup}

If we restrict the representation to the maximal compact subgroup $K=\{(0,0,\beta)\,,\beta\in[0,4\pi)\}= U^{(2)}(1)$, a double cover of $U(1)$, 
we obtain a unitary, although
reducible, representation of $U^{(2)}(1)$. It decomposes into $2j+1$ (for general $j$)
irreducible unitary characters, labeled by $m=-j,\ldots,j$.
This case coincides with that of a representation of $SU(2)$ restricted to $U^{(2)}(1)$. In a sense this case is trivial since it reduces
to a well-known, unitary case.

Denote by $\hat{A}_K$ the restriction of the resolution operator to the compact subgroup $K$ 
(as defined in eq. (\ref{AK})).  Since the representation restricted to 
$K$ is reducible, $\hat{A}_K$ is not proportional to the identity, and even non-invertible, in general. 
In this case  $\hat{A}_K$ is invertible only if the component-wise 
product of the vectors $\Phi$  and $\tilde{\Psi}$ is a vector with all its components different from zero. If all the components of this vector
are identical, then  $\hat{A}_K$ is proportional to the identity. 

The overlapping kernel restricted to $K$, $\kappa_K(k',k)$, is square-integrable on $U^{(2)}(1)$. Since in this
finite-dimensional case ${\cal V}$ and ${\cal V}^*$ are isomorphic, we can choose $\tilde{\Psi}=\Phi$
as before and therefore $\kappa_K(k',k)$ is positive definite and hermitian. Thus, $\kappa_K(k',k)$
is a reproducing kernel for the Hilbert subspace $Im(\hat{T})\subset L^2(U^{(2)}(1))$.

This case corresponds to taking $\gamma(z)=0$ in eq. (\ref{Hamiltonian}) (with the changes given in eq. (\ref{change})), leading to a unitary propagator identical to the one used for $SU(2)$
(known as $J_x$ array when $\lambda$ is constant \cite{JxArrayOptics,JxArrayOptics2}). The propagator produces usual $SU(2)$ coherent states (restricted to $U^{(2)}(1)$)
 starting from an arbitrary initial state.

\subsection{Restriction to a subgroup of elliptic elements}

Consider now the non-trivial case of a subgroup of elliptic elements in $SU(1,1)$ of the form $K_{[v]}=v K v^{-1}$, with
$v$ a representative element of the class $[v]\in SU(1,1)/U^{(2)}(1)\approx \mathbb{D}$ \cite{OPUC}. Restricting the representation to $K_{[v]}$ leads to a non-unitary finite-dimensional 
representation of $U^{(2)}(1)$, and this
corresponds (with the changes  (\ref{change})) to taking the propagator associated with a Hamiltonian in eq.
(\ref{Hamiltonian}) that is a compact operator. For the case of constant 
parameters $\gamma$ and $\lambda$, this corresponds to the case
$\gamma<\lambda$, see Fig. \ref{PTtrimer} (left). 
This propagator generates non-hermitian coherent states with 
bounded (but non constant) norm on $K_{[v]}$.

 The resolution operator is bounded and invertible for arbitrary nonzero vectors $\Phi$  and $\tilde{\Psi}$ (there are only a few values of $\nu$, depending
on the components of $\Phi$  and $\tilde{\Psi}$ for  which it is non-invertible). For a given non-zero $\Phi$ and a given $v$, there is a unique non-zero $\tilde{\Psi}$ (up to normalization) that makes 
the resolution operator proportional to the identity. The overlapping kernel is square-integrable on $K_{[v]}$.

\subsection{Restriction to subgroups of hyperbolic and parabolic elements}

The restriction to subgroups of parabolic or hyperbolic elements \cite{OPUC} always leads to non-square integrable representations for any choices of nonzero
$\Phi$ and $\tilde{\Psi}$.
The subgroup of parabolic elements correspond (with the changes  (\ref{change})) to the case $\gamma=\lambda$, see Fig. \ref{PTtrimer} (center), and subgroups of hyperbolic
elements correspond to the case $\gamma>\lambda$, see Fig. \ref{PTtrimer} (right).

In these last  cases
other strategies should be adopted to overcome non-square integrability, and this will be considered in the next subsection.

\subsection{Restriction to compact subsets}

The non square-integrability of the representation is caused by the integration on the non-compact subset $\mathbb{D}$. We can consider 
compact subsets of $\mathbb{D}$ (times $U^{(2)}(1)$) that avoid the non square-integrability. For instance, we can take the following subsets:

\begin{itemize}
 \item[a)] $\{\zeta,\zeta^*,\beta\}$, with $0<|\zeta|=r_0<1$ fixed. The integration in this case is over the compact subset 
$S^1\times U^{(2)}(1)$.

\item[b)]  $\{\zeta,\zeta^*,\beta\}$, with $0\leq r_m\leq |\zeta|\leq r_M<1$. The integration in this case is over the compact subset $[r_m,r_M]\times S^1\times U^{(2)}(1)$.

\item[c)]  $\{\zeta,\zeta^*,\beta\}$ with $\zeta=r_0e^{i \frac{2\pi k}{N}}$, $k=0,1,\ldots,N-1$, $0<r_0<1$. The integration in this
case is over $U^{(2)}(1)$ times a sum in $k=0,\ldots,N-1$.
\end{itemize}

In all these cases the resolution operator for the corresponding subfamily of coherent states is bounded and proper choices of
$\Phi$ and $\tilde{\Psi}$ can be made to render it invertible and even 
proportional to the identity. Case (c) is similar to the construction made in
\cite{samplingsphere, samplinghyperboloid,samplingHW} where
discrete (pseudo-) frames of coherent states were constructed for unitary
representations of $SU(2)$, $SU(1,1)$ and the Heisenberg-Weyl group,
respectively.

In none of these cases the subset constitutes a subgroup of $SU(1,1)$, not even an homogeneous space, i.e. of the form $SU(1,1)/H$ for some
subgroup $H$. Therefore some of the properties of Gilmore-Perelomov \cite{Perelomov} coherent states or coherent states modulo a subgroup 
\cite{CoherentStatesAAG} are lost. In particular, the subfamily of coherent states associated with 
them does not contain all the states generated by
the propagator for Hamiltonians with $\gamma\geq \lambda$, although they contain them for finite 
propagation distances (chosing appropriately $r_m$ and $r_M$ in case (b), for instance).

In any case, if the subfamily of coherent states associated with a compact subset leads to an invertible
resolution operator, they form an overcomplete family and any state of the system can be expressed in
terms of them.
%Also, the subfamily of coherent states is not preserved under the action of the propagator \ref{propagation}.

\section{Conclusions and outlook}

In this work coherent states associated with non-unitary representations of non-compact Lie groups have been introduced motivated by the behavior
of certain DPS with non-hermitian Hamiltonians (in particular $PT$-symmetric DPS), and following the general theory outlined in \cite{Zimmerman}. 
We have called these families of coherent states ``non-hermitian'' rather than ``non-unitary'' since we focus in the physical motivation 
(non-hermitian systems with a symmetry Lie group) rather than the mathematical setting (non-unitary representation of the symmetry Lie group).

Although we discuss for completeness
the general, infinite-dimensional case, which requires the use of Banach spaces (versus Hilbert spaces for the unitary case) to account
for the maximum generality, we shall focus  in the examples in finite-dimensional (infinite-dimensional examples will be considered elsewhere), non-unitary representations, where all the subtleties 
of Banach spaces do not appear. See \cite{Zimmerman} for infinite-dimensional examples with
non-trivial Banach spaces (${\cal V}=L^p(\mathbb{R})$ and ${\cal V}^*=L^q(\mathbb{R})$  with
$\frac{1}{p}+\frac{1}{q}=1$).

The main obstruction for constructing coherent states for non-unitary representations comes from
the lack of square-integrability of the representation. Different approaches have been used to circumvent this problem in the example 
of $SU(1,1)$ considered in Sec. \ref{exSU11}, the most interesting one being 
the restriction to certain subgroups (in particular elliptic subgroups), where the representation
restricted to the subgroup is still non-unitary but it is square-integrable.
This is related to the physically interesting case of a $PT$-symmetric waveguide array in the regime of unbroken $PT$-symmetry. Other possibilities
have been discussed, like the restriction to certain compact subsets of $SU(1,1)$.

Other approaches could be used to study this kind of systems.
For instance, we could use Riesz basis like in \cite{Bagarello1,Bagarello2,Bagarello3,Bagarello4,Bagarello5},
or the property of pseudo-hermiticity \cite{Mostafazadeh}.

% Bibliografia

\end{document}